\documentstyle[psfig]{mn}

\newcommand{\rmn}[1]{\mathrm {#1}}

\pagerange{\pageref{firstpage}--\pageref{lastpage}}

\pubyear{1996}

\begin{document}
\label{firstpage}

\title[The fundamental plane and dark haloes of ellipticals]
{On the fundamental properties of dynamically hot galaxies}
\author[A.G. Kritsuk]{Alexei G. Kritsuk$^{1,2}$\thanks{E-mail: \tt
agk@aispbu.spb.su}\\
$^1$ Institute of Astronomy, University of St
Petersburg, Stary Peterhof, St Petersburg 198904, Russia\\
$^2$ Max-Planck-Institut f\"ur Astrophysik, Postfach
1523, D-85740 Garching, Germany}

\date{Accepted 1996 August 2.
      Received 1996 April 24;
      in original form 1995 October 18}

\maketitle

\begin{abstract}
A two-component isothermal equilibrium model is applied to reproduce
basic structural properties of dynamically hot stellar systems
immersed in their massive dark haloes. The origin of the fundamental
plane relation for giant ellipticals is naturally explained as a
consequence of dynamical equilibrium in the context of the model. The
existence of two galactic families displaying different behaviour in
the luminosity--surface-brightness diagram is shown to be a result of a
smooth transition from dwarfs, dominated by dark matter near the centre, to
giants dominated by the luminous stellar component. The comparison of
empirical scaling relations with model predictions suggests that
probably a unique dissipative process was operating during the violent
stage of development of stellar systems in the dark haloes, and the
depth of the potential well controlled the observed luminosity of the
resulting galaxies. The interpretation also provides some restrictions
on the properties of dark haloes implied by the fundamental scaling
laws.   
\end{abstract}


\section{Introduction}
The term `dynamically hot galaxies' (DHGs) was introduced by Bender,
Burstein \& Faber \shortcite{bender..92} for a class of stellar
systems in which random motions provide most of the energy for support
of the system. This class includes all varieties of elliptical
galaxies from giant to dwarf and compact ellipticals, bulges of S0s
and spirals, as well as dwarf spheroidal (dSph) galaxies. 

High-quality
photometric and spectral data accumulated for a large number of objects
during the last decade, together with the application of more advanced
methods of statistical analysis, provided empirical scaling laws that
relate the global dynamical and structural observables of
DHGs, namely the central velocity dispersion of stars $\sigma_*$,
galaxy luminosity $L$, half-light radius $R_{\rmn{e}}$, and mean effective
surface brightness $I_{\rmn{e}}=L/2\pi R_{\rmn{e}}^2$
\cite[and references therein]{kormendy.89,djorgovski.90}. One of the
possible representations of such a law for a sample of giant
ellipticals (gEs) 
may be written as $L\propto \sigma_*^{3.45}I_{\rmn{e}}^{-0.86}$ 
\cite{djorgovski.87}. Biparametric relations of this kind known as the
fundamental plane (FP) indicate that the gEs populate a
surface in a three-space of observables. Various projections of the
surface reproduce correlations between the luminosity and
central velocity dispersion known as the Faber--Jackson
\shortcite{faber.76a} relation $L\propto\sigma_*^4$ (FJ) and between
the surface brightness and the effective radius
$\mu_{\rmn{e}}\simeq2.94\log{R_{\rmn{e}}}+const$ (HK), reported by Hamabe \&
Kormendy \shortcite{hamabe.87}. The bulges of early-type disc galaxies do
not obviously deviate from the HK parameter correlation for ellipticals
\cite{hamabe.87}. 

While compact ellipticals follow
nearly the same relations as the gEs, dwarf elliptical galaxies (dEs)
form a separate family characterized by distinct scaling laws. They
lie close to the FP for giant ellipticals but follow a different
luminosity--surface-brightness relation
$<\!\mu\!>_{\rmn{e}}\simeq0.75M_B+const$ \cite[hereafter BC]{binggeli.91}. As
far as the group of dwarf spheroidals is concerned, these galaxies also
follow the uniparametric BC relation for dEs but deviate from the FP
for gEs significantly (Bender et al. 1992; Burstein, Bender \& Faber 
1993)\nocite{bender..92,burstein..93}. 

The difference in the luminosity--surface-brightness relation for normal
and dwarf ellipticals is supplemented with
changes of galaxy surface brightness profiles along the galaxy
sequence from giants, most of which are well
described by the de Vaucouleurs $R^{1/4}$ empirical formula
(see, e.g., Burkert 1993\nocite{burkert93}), 
to diffuse dwarfs and dSphs, which are better fitted
by the exponential law \cite{faber.83}. These basic morphological
features must be reproduced by any successful physical model of galaxy
structure, formation and evolution. 

There is extensive literature on the modelling, comprehensively reviewed by
Ferguson \& Binggeli \shortcite{ferguson.94} and Gallagher \& Wyse
\shortcite{gallagher.94} with emphasis on dwarf galaxies, and by
Kormendy \& Djorgovski \shortcite{kormendy.89}, de Zeeuw \& Franx
\shortcite{de_zeeuw.91} and Burstein et al. \shortcite{burstein..93}
for giant ellipticals. It is generally suggested that the properties
of early-type galaxies are determined by dissipative collapse and then
modified by mergers. The details of cooling, star formation, and
feedback processes are considered to be important to produce the
luminosity--surface-brightness
relation (see references in the review papers above). However, each
particular physical model usually contains a large number of free input
parameters and therefore the interpretations do not appear to be unique.

One important source of uncertainties in the models is our poor
knowledge of the nature and distribution of dark matter in DHGs
\cite{kormendy88,ashman92,gallagher.94,de_zeeuw95}.
For spiral galaxies, flat H\,{\sc i} rotation curves suggest the
presence of extended isothermal dark haloes, and disc stability arguments
support the idea that the dark halo has a spheroidal shape
and, in any case, a much thicker distribution than the optical disc
\cite{bertin.93}. 
The detailed investigation of the main properties of the mass structure in
spirals resulted in derivation of the `universal rotation curve' -- a
counterpart to the FP of ellipticals (Persic, Salucci \& Stel
1996)\nocite{persic..96}. It shows two 
important features: (i) the distribution of dark matter in spirals is
completely different from that of the luminous matter, and (ii) the dark
matter content strongly and in a regular way depends on the luminosity, so
that low-luminosity galaxies have much more {\em dark} matter than luminous
matter, while luminous discs apparently dominate the internal dynamics of
very high-luminosity systems.
Until recently, neither observational nor theoretical arguments were
sufficient to provide crucial evidence for the presence of dark matter in
ellipticals \cite{bertin.93}. 
However, if ellipticals were formed without dark haloes, the
similarity between them and bulges of spirals would only be
superficial \cite{de_zeeuw.91}.

Combining the data on the inner ionized gas discs and on the H\,{\sc i}
discs extending to the outer regions of four elliptical galaxies, Bertola et
al. \shortcite{bertola...93} showed that the variation of the mass-to-light
ratio with galactocentric radius in ellipticals is qualitatively similar to that of
spirals. 
Recent measurements of the shape of the stellar line-of-sight velocity 
distribution out to two effective radii in four elliptical galaxies provide
definite evidence for the presence of massive dark haloes \cite{carollo....95}.
The mass profile for the elliptical galaxy NGC~4636, which is consistent with the stellar
velocity dispersion and with {\em ROSAT} and {\em ASCA} X-ray temperature
profiles,
suggests that the galaxy becomes dark matter dominated at roughly the de
Vaucouleurs radius \cite{mushotzky.....94}. A comparison of optical and
X-ray mass determinations of galaxies, groups, and clusters of galaxies
shows that most of the dark matter may reside in haloes around galaxies,
typically extending to $\sim200$ kpc for bright galaxies (Bahcall, Lubin \&
Dorman 1995)\nocite{bahcall..95}. The haloes may be stripped off in the
dense cluster environment to produce the suggested difference in
scaling relations for field and cluster ellipticals 
\cite{de_carvalho.92}. Unlike giant ellipticals, some of the dwarf
spheroidal galaxies have been found to show high {\em central}
mass-to-light ratios
\cite{aaronson.87,lake90a,mateo94,armandroff..95,vogt...95} that imply 
dSphs to be a class of dark matter dominated galaxies [note that there are 
alternative interpretations, e.g. Kuhn \& Miller \shortcite{kuhn.89}]. Thus the available data
suggest that the abovementioned features of dark matter distribution in
spirals may be relevant to the class of DHGs as well.

Obviously, the significant content of dark matter (either only
in the halo region or all over the stellar system) will control
the brightness profiles of the galaxies and therefore may play a major 
role in determining the observed global structural properties of
DHGs. The purpose of this paper is to check whether or not the fundamental
properties of DHGs can be explained solely as due to variations in
dark matter content and spatial distribution in the
galaxies. The preliminary answer is given here with the aid of an
equilibrium galaxy model, which includes the stellar component and the dark
component (both are isothermal, but with different temperatures) and
which suggests the existence of a `conspiracy' between the two following
a simple power-law dependence, cf. Burkert \shortcite{burkert94}. 
Such a two-component model allows one to 
reconcile the dissipative effects
(mainly related to the stellar, i.e. baryonic, component) 
with the dissipationless effects (related to the dark, presumably
collisionless component) in the development of galaxies. 
Thus it may provide some hints about the initial conditions and the nature
of star formation processes operating in DHGs.

The paper is organized as follows. Section 2 gives a detailed
description of the model. Section 3 discusses the FP for
giant ellipticals and the deviations of dSph galaxies from it
as they originate in the context of the model. In Section 4 
the distribution of DHGs in the luminosity--surface-brightness
diagram is compared with the model predictions. The results are
summarized in Section 5. 

\section{The two-component galaxy model}\label{GalMod}

Let $M(r)$ be a spherically symmetric configuration
of gravitating mass, including the stellar (luminous) component of a galaxy
and its isothermal dark halo. The equilibrium dark matter distribution
in a gravitational
potential $\phi(r)$ of $M(r)$ can be described by
\begin{equation}
\sigma_{\rmn{DM}}^2\frac{\rmn{d}\;
\ln{\rho_{\rmn{DM}}}}{\rmn{d}\:{r}} = -\nabla\phi\equiv -{G M(r)\over r^2}.
\end{equation}
The stellar density distribution of the galaxy, $\rho_*$, responding to the
same gravitational potential, satisfies
\begin{equation}
\sigma_{\rmn{*}}^2\frac{\rmn{d}\: \ln{\rho_*}}{\rmn{d}\:{r}} = -\nabla\phi,
\end{equation}
where $\sigma_{\rmn{DM}}$ and $\sigma_*$ are the one-dimensional velocity
dispersions of dark matter particles and stars (it is assumed that both do
not depend on the radius). The Poisson equation
\begin{equation}
\Delta\phi=4\pi G(\rho_{\rmn{DM}}+\rho_*)
\end{equation}
closes the equation set.
A set of four parameters ($\rho_{\rmn{DM},0}$,
$\rho_{*,0}$, $\sigma_{\rmn{DM}}$, and $\sigma_*$) completely defines
the model. 

It is straightforward to reduce the 
order of the system since the `hydrostatic' dark matter and stellar
densities are related via 
\begin{equation}
\frac{\rmn{d}\: \ln{\rho_{\rmn{DM}}}}{\rmn{d}\:\ln{\rho_*}} = {\sigma^2_*\over
\sigma_{\rmn{DM}}^2} \equiv\beta,\label{Beta}
\end{equation}
or
\begin{equation}
\rho_{\rmn{DM}}={\cal C}\rho_*^{\beta},\label{GasDenPowerLaw}
\end{equation}
where ${\cal C}$ is the integration constant and $\beta$ is the ratio of specific
kinetic energies of stars and dark matter particles.

Given equation (\ref{GasDenPowerLaw}), one can determine the total mass 
distribution using the modified Poisson equation
\begin{equation}
{1\over r^2}{\rmn{d}\over \rmn{d}r}\left(r^2{\rmn{d}\ln\rho_*\over
\rmn{d}r}\right)=-{4\pi
G\over \sigma_*^2}(\rho_*+\rho_{\rmn{DM}}). \label{Poi}
\end{equation}
Substituting dimensionless variables
\begin{equation}
x=r/r_0,
\end{equation}
\begin{equation}
y=\rho_*/\rho_{*,0},
\end{equation}
where
\begin{equation}
r_0=\frac{\sigma_*}{\sqrt{4\pi G \rho_{*,0}}}, \label{king_rad}
\end{equation}
and introducing a new parameter as the ratio of the densities at the
centre, 
\begin{equation}
\delta=\rho_{\rmn{DM},0}/\rho_{*,0},
\end{equation}
one gets a generalized equation of Emden-Fowler type:
\begin{equation}
{1\over x^2}{\rmn{d}\over \rmn{d}x}\left(x^2{\rmn{d}\ln y\over
\rmn{d}x}\right)=-(y+\delta y^{\beta}).\label{major}
\end{equation}
It yields a biparametric family of solutions $y(x;\beta,\delta)$ for the 
initial conditions $y(0)=1$, $y^{\prime}(0)=0$. 

\begin{figure*}
\psfig{figure=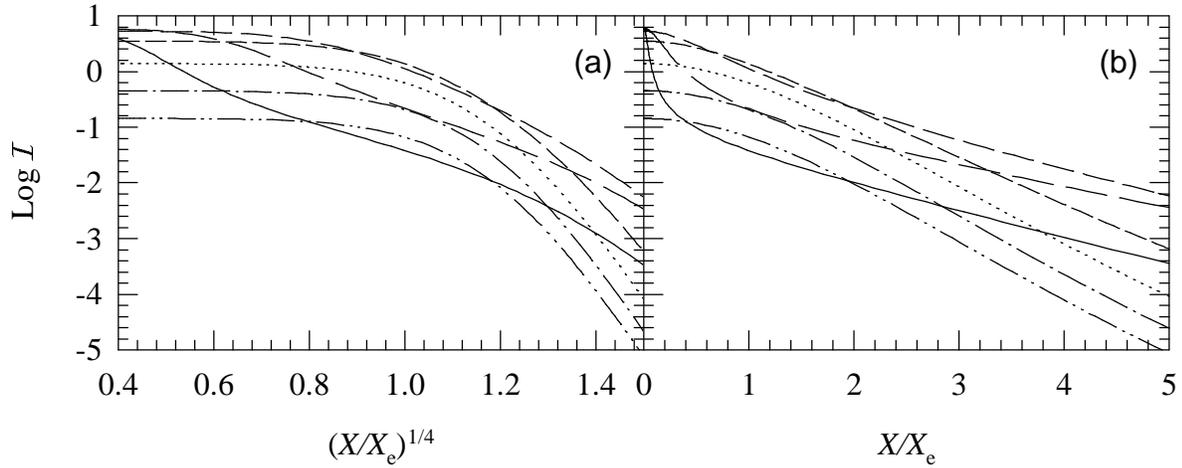,clip=}
\vspace{-21.9cm}
\caption{Simulated surface brightness profiles for $\beta=0.2$. The sequence
of line styles: solid, dashed (long, medium, short), dotted, dashed
dotted, dashed
double dotted corresponds to $\delta=0.001, 0.01, 0.1, 1, 10, 100, 1000$,
respectively. Panels (a) and (b), having different horizontal scales,
display $R^{1/4}$ and exponential profiles as straight lines; radial
distance $X$ is given in units of the half-light radius $X_{\rmn{e}}$, see
equation (15)
%(\ref{erad})
below. The surface brightness ${\cal I}$ is given in dimensionless units defined
by equation (17).}
\label{prfl02}
\end{figure*}
\begin{figure*}
\psfig{figure=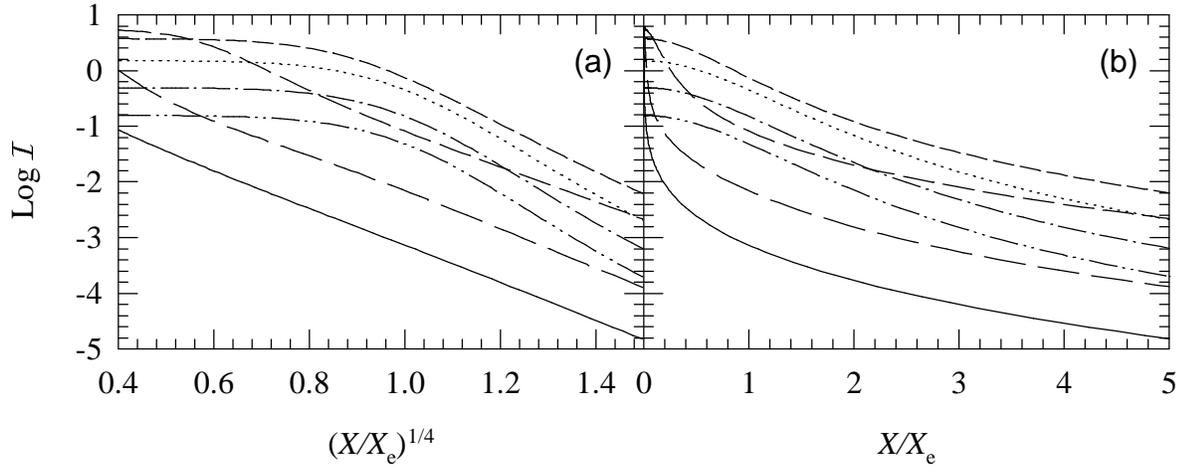,clip=}
\vspace{-21.9cm}
\caption{As Fig. 1 but for $\beta=0.5$. }
\label{prfl05}
\end{figure*}
\begin{figure*}
\psfig{figure=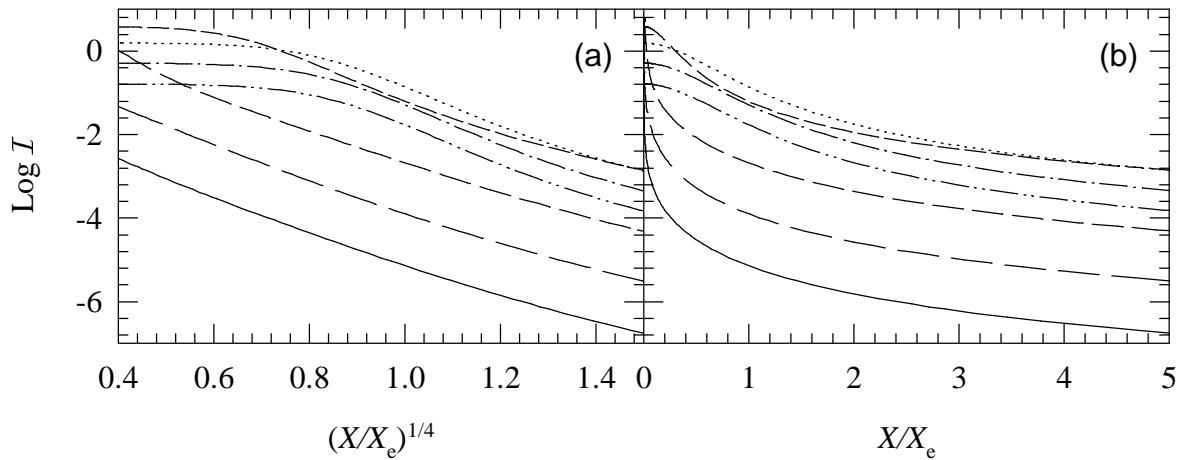,clip=}
\vspace{-21.9cm}
\caption{As Fig. 1 but for $\beta=0.6$. Please note the
difference in the ordinate scale compared with Figs 1 and 2. }
\label{prfl06}
\end{figure*}

Now each particular model galaxy can be characterized by a pair of
dimensional parameters (e.g., $\sigma_*$ and $r_0$) that control the scaling,
and a pair of dimensionless parameters $\beta$ and $\delta$ that determine
the shape of the light distribution in the galaxy. 
In this framework the structural homology generally assumed for the class of
E galaxies implies the constancy of $\beta$ and $\delta$.
Since there are indications that the homology is broken (see, e.g., 
Djorgovski 1995\nocite{djorgovski95}; Hjorth \& Madsen
1995\nocite{hjorth.95}), all four parameters are treated as free in 
the following. 

Parameter $\beta$ defines the so-called `conspiracy' of the dark and
luminous matter. 
In the case of $\beta=1$ the model describes a self-gravitating isothermal
sphere of infinite mass with uniform mass-to-light ratio. A less concentrated dark
matter distribution requires $\beta\in[0,\; 1)$. Asymptotically
$\rho_*\sim r^{-2/\beta}$ when 
$r\rightarrow\infty$, and the mass of the stellar system is finite
only for $\beta<2/3$ if no tidal cut-off is introduced.

Parameter $\delta$ controls the dominant mass component in the centre of a
galaxy, i.e. the galactic central mass-to-light ratio:
\begin{equation}
(M/L)_0={\rho_{*,0}+\rho_{\rmn{DM},0}\over\rho_{*,0}(M/L)_*^{-1}}=
(1+\delta)(M/L)_*;\label{ml}
\end{equation}
here $(M/L)_*$ is the mass-to-light ratio of individual stars. Varying
$\delta$ in the range from zero to $\sim100$ one can cover the wide
interval of $M/L$ estimates based on observations of DHGs.

When $\beta=1/2$ and $\delta\la0.1$, the stellar system has a density
distribution of an isothermal sphere in the inner region and a cut-off
at the core radius of the isothermal dark halo. In this case the
density of luminous material varies as $r^{-2}$ in the intermediate
range of $r$ outside the core region and asymptotically as $r^{-4}$ in the
outer region where the dark component dominates by mass. It is
this behaviour that is required by the de Vaucouleurs surface brightness
profile \cite{bertin.93}. %page 515 
Figs \ref{prfl02}--\ref{prfl06} illustrate the variety of profiles
responding to  different values of $\delta$ and $\beta=0.2, 0.5,
0.6$. While for $\delta\la0.1$ the 
profiles closely follow the empirical $R^{1/4}$ law in a range of radii
$0.1R_{\rmn{e}}\le r\le1.5R_{\rmn{e}}$ (Burkert 1993, 1994)\nocite{burkert93,burkert94}, they
look more like exponential at $\delta\ga10$. The farther $\beta$
deviates from 1/2, the worse fit the de Vaucouleurs formula yields in the
limit of small $\delta$. As a result the profiles generated for $\beta=0.2$
or 0.6 do not satisfy the observational constraints on the quality of fit
with the de Vaucouleurs formula established by Burkert \shortcite{burkert93}
who carried out a systematic analysis of CCD data for a large sample of
E galaxies. 

The physical meaning of the condition $\beta=1/2$ can be readily understood
in terms of a tentative dissipative galaxy formation scenario. 
When baryonic gas is initially thermalized in the dark halo potential well,
its specific thermal energy equals the energy of dark matter particles and
its density distribution follows that of the dark matter. If the efficiency
of subsequent star formation follows the Schmidt law
$\dot\rho_*\propto\rho^2_{\rmn{gas}}$, the stellar density distribution will
automatically satisfy the $\beta=1/2$ condition. 
As the observed light profiles of gEs show a preference for such a choice
of $\beta$, it will be considered as a working hypothesis hereafter.

\begin{figure}
\psfig{figure=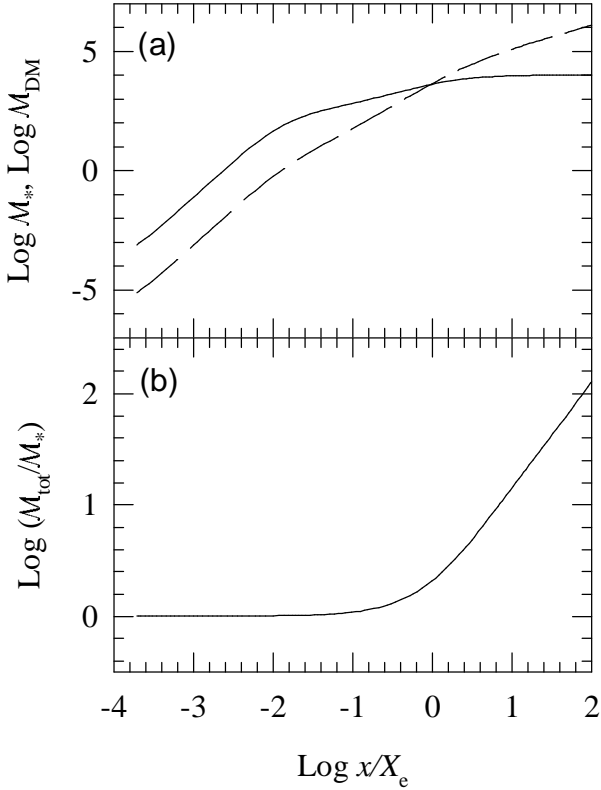,clip=}
\vspace{-16.9cm}
\caption{Mass of the luminous (solid line) and dark (dashed line) matter
(panel {\bf a}) and integrated mass-to-light ratio (panel {\bf b})
versus radius for the model with $\beta=0.5$ and $\delta=0.01$. }
\label{mass05-2}
\end{figure}

\begin{figure}
\psfig{figure=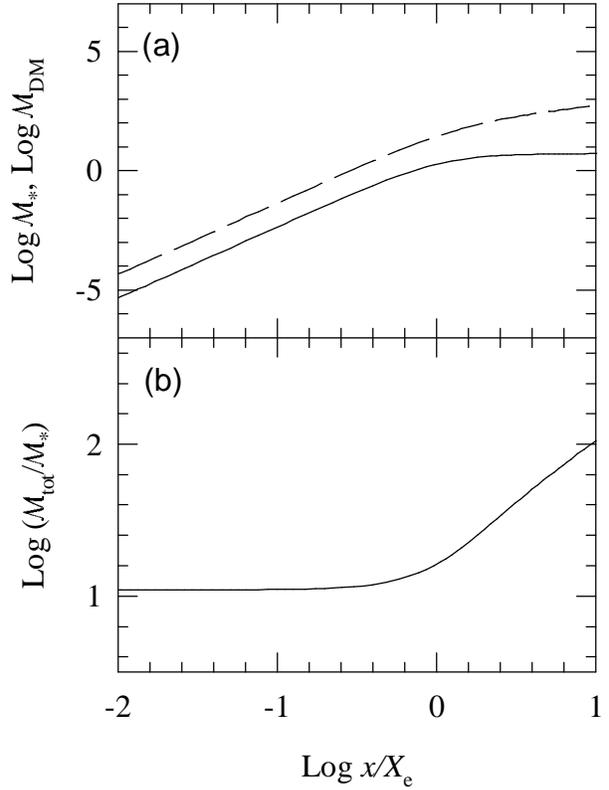,clip=}
\vspace{-16.9cm}
\caption{As in Fig. 4 but for $\delta=10$. }
\label{mass05_1}
\end{figure}

The distributions of dark and luminous mass for $\delta=0.01, 10$ and
$\beta=1/2$, typical for the model galaxies with (conventionally)
`de Vaucouleurs' and `exponential' profiles, are shown in Figs 
\ref{mass05-2}(a) and \ref{mass05_1}(a). 
The mass of the stellar system is defined here as
\begin{equation}
{\cal M}_*(x)\equiv4\pi\int_0^{x}x^2y\rmn{d}x.
\label{mass}
\end{equation}
The ratio of the total (dark$+$luminous) mass to ${\cal M}_*$ is given
in Figs \ref{mass05-2}(b) and \ref{mass05_1}(b) 
to illustrate how the integral mass-to-light ratio depends on the radius.
Note that when $\delta=0.01$ the masses of stars and dark matter
enclosed inside $R_{\rmn{e}}$ are comparable (cf. Saglia, Bertin \&
Stiavelli 1992)\nocite{saglia..92}, but outside the half-light radius hidden 
matter dominates by mass that diverges $\sim r$ when $r\rightarrow\infty$. 
This inappropriate asymptotic behaviour does not significantly modify the
global photometric quantities of the model galaxies, such as the luminosity
or the mean effective surface brightness, while the value of $\beta$ is not
too close to 2/3. 
However, even for $\beta=0.5$ and $\delta\la0.01$ the adopted isothermal
approximation may result in a systematic overestimate of the luminosity by a
factor of $\sim1.5$, since the model predicts nearly perfect $R^{1/4}$ light
profiles with no cut-off for radii up to $>5R_{\rmn{e}}$ (see Fig.
\ref{prfl05}), i.e. far outside the region typically covered by observations. 
The two-component self-consistent models of Bertin, Saglia \&
Stiaveli \shortcite{bertin..92}, that take into account the effects of
anisotropies expected to be present in both components, would be a
better representation for real galaxies.
Still, the isothermal model may be used as a zeroth-order approximation, since
it is able to reproduce the general trend in global structural properties
over the sequence of DHGs from giants to diffuse dwarf
ellipticals.\footnote{This model does not account for cluster-related 
phenomena, like the extended shells around cDs, which are beyond the scope
of this paper.} The model suggests that the central mass-to-light
ratio may indeed be the major parameter controlling the shape of light
profiles of the galaxies. 
In order to find a range of realistic values of $\beta$ and $\delta$, one
should simulate the distribution of model galaxies in the 3D space of
observables. This will be the subject of the following sections. 

\section{An edge-on view of the fundamental plane}
The fundamental plane for ellipticals is argued to contain clues to the
initial conditions and processes of galaxy formation, being not just a
consequence of the virial theorem.
A simple reason behind this statement is that scale-free pure dynamics
cannot explain the origin of the dimensional scaling laws \cite{bertin.93}.
If the existence of dark haloes is allowed, the homology hypothesis
appears to be violated. Then a `conspiracy' between the luminous and dark
matter is able to introduce fractional power-law indices into the scale-free
dynamical relationships and thereby into some of the observed correlations. 
This issue will be illustrated here with the aid of the two-component
isothermal model.


Let ${\cal L}$, $X_{\rmn{e}}$, and ${\cal I}_{\rmn{e}}$ be 
{\em dimensionless}\, 
luminosity, half-light radius, and mean effective surface brightness,
respectively, so that 
\begin{equation}
{\cal L}\equiv{\cal M}_*(\infty)= {4\pi G(M/L)_*\over r_0 \sigma^2_*}L,
\label{ldless}
\end{equation}
\begin{equation}
{\cal L}/2=2\pi\int_0^{X_{\rmn{e}}}{\cal I}(X)X\rmn{d}X, \label{erad}
\end{equation}
\begin{equation}
{\cal I}_{\rmn{e}}={{\cal L}\over2\pi X_{\rmn{e}}^2},
\end{equation}
where
\begin{equation}
{\cal I}(X)\equiv2\int_X^{\infty}{yx\rmn{d}x\over\sqrt{x^2-X^2}}=
{4\pi G r_0 (M/L)_*\over\sigma^2_*}I(R)
\end{equation}
is the dimensionless surface brightness and $X=R/r_0$ is the radial
distance in projection. 

These integral quantities can be readily calculated as functions of $\beta$
and $\delta$.
In particular, for $\delta\gg1$ the second term dominates in the {\em rhs} of
equation (\ref{major}), so the general solution can be approximated by the
rescaled $y(\delta=1)$ solution:
$y(x; \beta, \delta)\approx y(x/\sqrt{\delta}; \beta, 1)$.
The integral characteristics can be expressed as functions of $\delta$ 
explicitly: 
\begin{equation}
{\cal I}(x,\delta)\approx{{\cal I}(x,1)\over\sqrt{\delta}},\label{sub}
\end{equation}
\begin{equation}
X_{\rmn{e}}(\delta)\approx {X_{\rmn{e}}(1)\over\sqrt{\delta}}, 
\end{equation}
and
\begin{equation}
{\cal L}(\delta)\approx\delta^{-{3\over2}}{\cal L}(1),\label{lum}
\end{equation}
where $\delta\gg1$. At the other extreme, $\delta\la1$, there is no such
simple analytical approximation owing to a more complicated restructuring of
solutions, so a numerical technique has to be used.

In order to study the structural properties of dynamically hot
galaxies, Bender et al. \shortcite{bender..92} have defined an
orthogonal coordinate system, which is termed as the $\kappa$-space: 
\begin{equation}
\kappa_1\equiv(\log{\sigma_0^2}+\log{R_{\rmn{e}}})/\sqrt{2},
\end{equation}
\begin{equation}
\kappa_2\equiv(\log{\sigma_0^2}+2\log{I_{\rmn{e}}}-\log{R_{\rmn{e}}})/\sqrt{6},
\end{equation}
\begin{equation}
\kappa_3\equiv(\log{\sigma_0^2}-\log{I_{\rmn{e}}}-\log{R_{\rmn{e}}})/\sqrt{3}.
\end{equation}
This coordinate system based on observables has been used to interpret the
properties of DHGs in terms of known physical processes. The projection of
galaxy distribution on the plane $\kappa_3=const$ shows the FP close to face-on. 
In contrast, the $\kappa_1=const$ and $\kappa_2=const$ projections show two
edge-on views of the FP. 
A major conclusion of Bender et al. \shortcite{bender..92} was that all
types of DHGs except the extreme dSphs lie nearly in the same FP as defined 
by the giant ellipticals. 

\begin{figure}
\psfig{figure=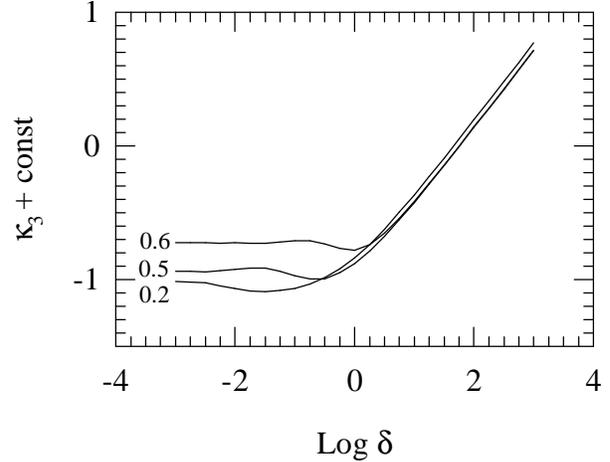,clip=}
\vspace{-23.2cm}
\caption{Plot of $\kappa_3$ versus $\delta$ for the grid
of models with $\beta=0.2, 0.5, 0.6$.}
\label{kappa3}
\end{figure}

While $\kappa_1$ and $\kappa_2$ coordinates are
essentially dimensional, $\kappa_3$ is proportional to the central
mass-to-light ratio (as follows from King's core-fitting formula)
and can be expressed as a combination of the
dimensionless variables defined above:
\begin{equation}
\kappa_3=(\log{X_{\rmn{e}}}-\log{\cal L}+const)/\sqrt{3}.
\end{equation}
This allows for identification of a scale-free counterpart to the edge-on
projection of the FP in terms of dimensionless variables
$({\cal I}_{\rmn{e}}, \cal L, \delta)$, which is shown in Fig.
\ref{kappa3}.
As the realistic range for $\delta$ is undefined at this
stage, the graphs are given for a wide interval from $10^{-3}$ to $10^3$;
also for completeness three cases of $\beta$ are shown.
The asymptotics $\kappa_3\sim const$ for $\delta\ll1$ (the virial limit) and
$\kappa_3\sim(\log{\delta}+const)/\sqrt{3}$ for $\delta\gg1$ [see the
approximations (\ref{sub})--(\ref{lum})] are common to all
represented values of $\beta$.
In the intermediate range of $\delta$ ($-1.5<\log{\delta}<0.5$ for
$\beta=0.5$), where the solution undergoes a transition from one asymptotic
behaviour to another, one can see a slight negative tilt ($\Delta\kappa_3\simeq-0.1$) for
$-1.5<\log{\delta}<-0.2$, a minimum at $\delta\simeq-0.2$ and a rise at
higher $\delta$.
These changes are implied by the non-trivial restructuring of the model
galaxies in the course of the transition from luminous to dark matter
dominated systems.
If there is a one-to-one mapping between the luminosity (or `the mass in the
luminous confines' $\kappa_1$\footnote{Note that $\kappa_1$ was defined to
represent roughly the mass of the galaxy `within the luminous confines',
assuming no rotational support and no structural variations among the
galaxies.}) and the dark matter content $\delta$ of DHGs, a comparison of the
galaxy distribution in $(\kappa_1,\kappa_3)$ projection [see fig. 1 from
Burstein et al. \shortcite{burstein..93}] and the $\kappa_3-\log{\delta}$
relation makes sense. 
The FP of gEs can be identified with a slightly tilted part of the
$\beta=0.5$ curve (cf. Renzini \& Ciotti 1993\nocite{renzini.93}; Ciotti,
Lanzoni \& Renzini 1996\nocite{ciotti..96}).
Dwarf ellipticals, having progressively higher dark matter content, show
small positive deviations from the FP.
Finally, the dark matter dominated dSphs are characterized by extremely
high mass-to-light ratios. 
If $\beta\simeq0.5$ the deviations from the FP for gEs (roughly
$\Delta\kappa_3\simeq 1.1$ dex) demonstrated by the two most extreme dSph
galaxies from the sample used in Burstein et al. \shortcite {burstein..93}
correspond to $\delta\simeq32$. 
Equation (\ref{ml}) then gives a central mass-to-light ratio of about
$33(M/L)_*$, which is close to the values reported for Draco and 
Ursa Minor \cite{aaronson.87,lake90a,armandroff..95}. 

Thus the scale-free dependences implied by the two-component model are
able to explain the gross features of edge-on projections for the
distribution of DHGs about the FP for giant ellipticals. 
Similar analysis for the face-on view of the FP, which cannot be reduced to
scale-free considerations, will be a subject of the following section.  

\section{The synthetic luminosity--surface-brightness diagram}
Within the FP, dynamically hot galaxies separate into two apparent
sequences \cite{binggeli..84,kormendy85}, which are clearly seen in the 
luminosity--surface-brightness diagram (see, e.g., Capaccioli, Caon \&
D'Onofrio 1992, 1993)\nocite{capaccioli..92,capaccioli..93}. 
The main sequence is defined by normal ellipticals (giant and intermediate),
bulges, and compact ellipticals. 
Along this sequence, surface brightness decreases with growing luminosity. The
second sequence is defined by dwarf ellipticals (nucleated and diffuse) and
dwarf spheroidals \cite{bender..92,burstein..93}. 
The surface brightness grows with luminosity on this branch. 
A handful of compact ellipticals continue the main sequence towards
lower luminosities beyond the turning point to the dwarf 
branch.\footnote{This group of galaxies is assumed
to have a specific formation history which controls their structural
properties (see Burkert \& Truran 1994)\nocite{burkert.94}, and therefore it
stays outside the scope of this paper.}
\begin{figure}
\psfig{figure=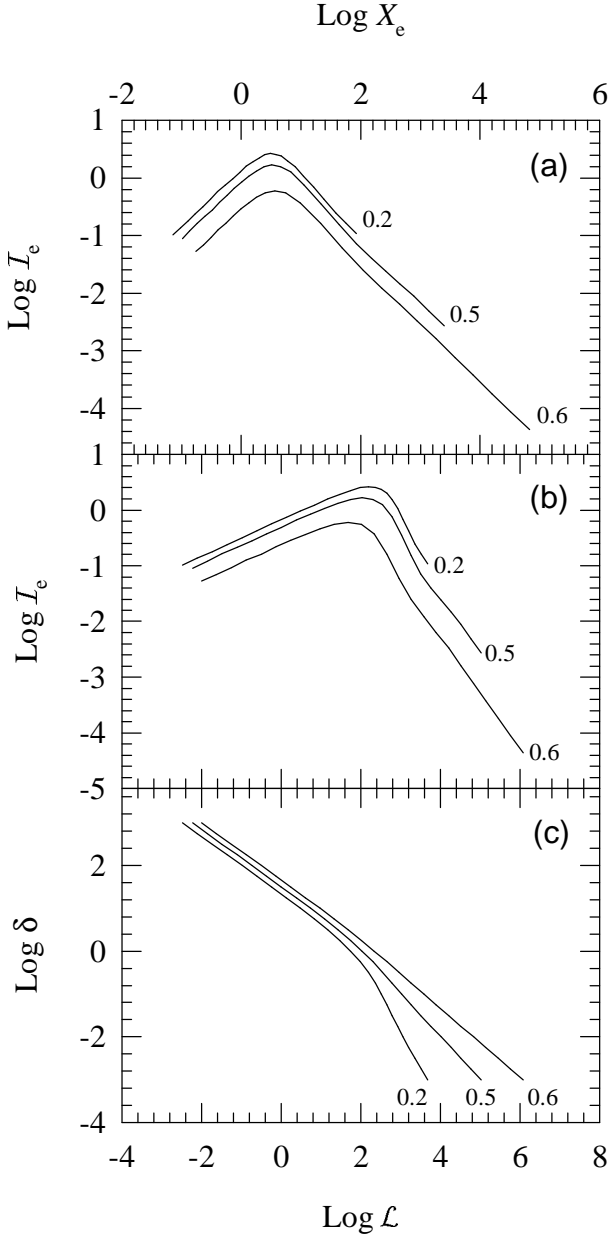,clip=}
\vspace{-11.3cm}
\caption{Global photometric quantities and the central
dark-to-luminous mass density ratios for the set
of models with $10^{-3}<\delta<10^3$ and $\beta=0.2, 0.5, 0.6$. (a) Mean effective
surface brightness versus effective radius. (b) Mean effective
surface brightness versus luminosity. (c) Central dark matter fraction $\delta$
versus luminosity.  Note that axes are scaled in dimensionless units.}
\label{diagram}
\end{figure}

The scale-free counterparts for the distribution of galaxies in the
effective-radius--surface-brightness and luminosity--surface-brightness
planes are shown in Figs \ref{diagram}(a) and (b).
Plots of $\log{{\cal I}_{\rmn{e}}}$ versus $\log{X_{\rmn{e}}}$ and versus
$\log{\cal L}$ are given for a grid of models with $\beta=0.2, 0.5, 0.6$ and
$\delta$ varying from $10^{-3}$ to $10^3$. 
The curves $\beta=const$ display a characteristic shape of the empirical
$R_{\rmn{e}}-<\!\mu\!>_{\rmn{e}}$ and $M_B-<\!\mu\!>_{\rmn{e}}$  diagrams
\cite[fig. 2]{capaccioli..93}, changing the sign of their slopes at
$\delta\simeq1$. However, a direct comparison of these scale-free diagrams
with those based on observables is not possible since
$L\propto r_0\sigma^2_*{\cal L}$,
$I_{\rmn{e}}\propto r_0^{-1}\sigma^2_*{\cal I}_{\rmn{e}}$, and
$R_{\rmn{e}}\propto r_0 X_{\rmn{e}}$, while $r_0$ and $\sigma_*$ are
unknown functions of $\delta$.
The major task of this section is to determine these hidden parametric
relationships (or their equivalents) and to delimit the range of
realistic values of $\delta$ for the giant and dwarf branches separately.

Our considerations will be based on the following assumptions:
\begin{enumerate}
\item
the value of $\beta$ is unique for both `giant' and `dwarf'
families of galaxies;
\item$\beta\simeq0.5$ since only in this case the
model reproduces the $R^{1/4}$ brightness profiles of normal
ellipticals;
\item $\delta$ is the key quantity which traces the position of any
particular galaxy in the three-space of observables;
\item
the intersection point of the dwarf and giant branches
coincides with a maximum in ${\cal I}_{\rmn{e}}(\cal L)$ dependence.
\end{enumerate}
This most straightforward approach to comparing the model with
observations is not the only possibility. In particular, giant
ellipticals are known to represent a biparametric family of galaxies
and perhaps it cannot be described by varying only one dimensionless
model parameter. However, the insights provided by the model are
considered here only as indicating the direction for a more detailed 
study. Therefore, following the simplicity concept, the interpretation
starts from the assumptions above. 

Using power-law approximations for the observational scaling relations:
\begin{equation}
L\propto I_{\rmn{e}}^A\sigma_*^{2B},
\end{equation}
\begin{equation}
L\propto \sigma_*^{2C},
\end{equation}
and for the dependencies between the dimensionless quantities provided
by the model (see Figs \ref{diagram}b and c):
\begin{equation}
{\cal I}_{\rmn{e}}\propto {\cal L}^D,
\end{equation}
\begin{equation}
\delta\propto {\cal L}^E,
\end{equation}
after some algebra one can show that if the model fits the observational
relations then
\begin{equation}
\delta\propto L^F,
\end{equation}
where
\begin{equation}
F={E(C+AC-B-2A)\over AC(D+1)}.\label{key}
\end{equation}
The estimates for index values are given in Table 1. When
$F$ is known, the $\delta$-scale can be calibrated in the 
luminosity units, which allows one to get rough estimates of the central
dark-to-luminous mass density ratio for galaxies with known
luminosity.  
\begin{table}
\caption{Power-law coefficients for the observed and simulated scaling
relations.}  
\begin{tabular}{@{}cllll}
Index&Giants&&Dwarfs\\\hline
$A$ & $-0.86^1$		&       	& $\;\;\:1.33^2$        	\\
$B$ & $\;\;\:1.73^1$	&		& $\;\;\:0$			\\
$C$ & $\;\;\:2.0^3$	& $\;\;\:1.2^4$	& $\;\;\:2.0$	& $\;\;\:2.8^5$	\\
$D$ & $-1.3^6$		& $\;\;\:0.17^7$& $\;\;\:0.31^8$& $\;\;\:0.31^8$\\
$E$ & $-1.0^6$		& $-0.8^7$	& $-0.7^8$	& $-0.7^8$	\\
$F$ & $-0.52$		& $-0.06$	& $-0.40$	& $-0.55$	\\
$H$ & $-1.04$		& $-11.1$	& $-1.6$	& $-1.2$	\\\hline\\\end{tabular}

\medskip

$^1$ The FP for giant ellipticals \cite{djorgovski.87}.\\
$^2$ The luminosity--surface-brightness relation for Virgo dwarfs
 \cite{binggeli.91}.\\
$^3$ The Faber--Jackson \shortcite{faber.76a} relation.\\
$^4$ The luminosity--velocity-dispersion relation for dwarfs \cite{davies....83}.\\
$^5$ As note 4 \cite{peterson.93}.\\
$^6$ The approximation for $\beta=0.5$ and
$\log{\delta}\in[-1.5,\:-0.5]$.\\
$^7$ As note 6, for $\log{\delta}\in[0,\:0.5]$.\\
$^8$ As note 6, for  $\log{\delta}\in[0,\:3]$.\\
\end{table}
With $F=-0.52$ the range of absolute magnitudes of gEs,
$-25.5<M_B<-19.5$ \cite{capaccioli..93}, corresponds to a
difference of $\sim1.3\;\rmn{dex}$ in $\delta$. The intersection point
of two galaxy branches in the $({\cal L}, {\cal I}_{\rmn{e}})$ plane is located
at $\log{\delta}\simeq-0.2$. Thus the brightest galaxies may have
$\log{\delta}\simeq-1.5$, which means that they contain $\sim30$ times
more mass in stars than in dark matter in their core
regions. At the same time, this range of $\delta$ confirms the
suggested identification of the FP for gEs (see Section 3).

Note 
that a part of the mass that is conventionally termed here as `dark'
may be present in gEs as a hot gas emitting X-rays. The equilibrium
density of this hot gas follows the dark matter distribution, and
therefore fits the same model for the gravitational potential (with
$\rho_{\rmn{gas}}+\rho_{\rmn{DM}}$ substituted for
$\rho_{\rmn{DM}}$). The typical ratio of the central stellar and hot gas
densities for giant ellipticals estimated on the basis of the hydrostatic
equilibrium model is $\sim10^4$, so its contribution
to $\delta$ must be negligible \cite{kritsuk96}. 

For dwarf galaxies the major source of uncertainty is the poorly
known luminosity--velocity-dispersion relation [cf. Held et al.
\shortcite{held...92} and Peterson \& Caldwell \shortcite{peterson.93}], so
three cases are represented 
in Table~1: the flat correlation $L\propto \sigma_*^{2.4\pm0.9}$ found by Davies et
al. \shortcite{davies....83}, the much steeper dependence
$L\propto \sigma_*^{5.6\pm0.9}$ suggested by Peterson \& Caldwell
\shortcite{peterson.93}, and the luminosity--velocity-dispersion relation
with the original FJ index. 
The observed range of luminosities of 4.6 dex from $M_B=-8$ to
$M_B=-19.5$ implies the value of $\delta$ at the faint end of
this branch to be roughly 2, 260, 60 for the three index values, respectively. 
Thus the normal FJ law gives an estimate of $\delta$ which is in rough
agreement with the one determined in \S3 for the extreme
dSph galaxies of $M_V\simeq-9$. 

The implemented fitting procedure also yields the index value $H$ for the
relation between the central stellar density and $\delta$: 
\begin{equation}
\rho_{*,0}\propto\delta^H,
\end{equation}
\begin{equation}
H={(2C-A-3AD-2B-2ACD)\over E(C+AC-B-2A)}
\end{equation}
(see Table 1). The value of $H\approx-1$ obtained for the giant branch
implies a weak dependence of $\rho_{\rmn{DM,0}}$ on $\delta$ along the
gE sequence, cf. Kormendy \shortcite{kormendy88}. In case of dwarf
galaxies the uncertainties in $F$ and 
$H$ values are high due to the large spread of points in the
$(L, I_{\rmn{e}})$ plane and the poorly determined value of $C$, but
specifically steep luminosity--velocity-dispersion relations would
roughly reproduce the scalings predicted by Dekel \& Silk
\shortcite{dekel.86} for their dark halo dominated model. 

Undoubtedly, further observations are 
required to clarify the situation and only preliminary conclusions can
be drawn at this stage, but it is not ruled out by the
above analysis of the empirical scaling laws that both giant and dwarf
galactic sequences are characterized by the same values of $F$ and $H$.

In a special case when $F=-0.5$ and $H=-1$ the
central stellar density and velocity dispersion would be related via
$\rho_{\rmn{*,0}}\propto\sigma_*^2$ and the characteristic radius
$r_0$ (which is proportional to the so-called King radius)
would be a constant (see equation \ref{king_rad}). Moreover, since it is
assumed that $\beta=1/2$, it follows from equations (\ref{Beta}) and
(\ref{GasDenPowerLaw}) that 
$\rho_{\rmn{*}}(r)\propto\rho^2_{\rmn{DM}}(r)$ and 
$\sigma^2_*=0.5\sigma^2_{\rmn{DM}}$. Therefore the
distribution of stellar density in galaxies may be written as
\begin{equation}
\rho_{\rmn{*}}(r)=\left[8\pi G r_0^2\rho^2_{\rmn{DM,0}}\right]^{-1}
\sigma_{\rmn{DM}}^2\rho^2_{\rmn{DM}}(r),\label{out}
\end{equation}
where the expression in square brackets is a constant. Since
$\sigma^2_{\rmn{DM}}$ is related to the depth of the potential well,
equation (\ref{out}) implies that the initial burst of star formation (from
baryonic gas thermalized in a potential of the dark halo) is regulated by a
dissipative binary collisional process and the mass of the resulting
stellar system is restricted (via the gradually increasing feedback
heating) by the depth of the potential well. The final quasi-steady
state which is observed can be reached either due to gas repulsion
(winds driven by supernovae in the case of galaxies at the dwarf end) or
due to gas consumption (the case of the giant end), and a combination of both may operate during the
formation of intermediate ellipticals and bulges. Thus the ability to
process baryonic gas into stars, controlled by the properties of dark
haloes, provides the essential morphological features of DHGs. 

This interpretation can hint at a coherent theory of
dissipative galaxy formation, which must be able to reproduce the
mass--luminosity-density--metallicity relation
\cite{djorgovski.90,matteucci92}. The approach to such a 
theoretical scheme, however, requires details which must be considered
elsewhere. 

\section{Conclusions}
A two-component spherically symmetric galaxy model with a stellar system
immersed in an isothermal massive dark halo has been applied to
reconstruct basic properties of dynamically hot galaxies. While being
obviously too crude to describe the observed detailed photometric and
kinematical features (like isophote ellipticity and twisting,
boxy and discy shapes, dynamically distinct dense nuclei, ripples and
shells, etc.), the model reproduces the essence of changes in
galaxy structure along the morphological sequence dSph--dE--gE 
(including bulges of S0 and spiral galaxies). 

The conclusions of this study can be formulated as follows.
\begin{enumerate}
\item Dark haloes of dynamically hot galaxies can play an important role in
controlling the shape of their surface brightness profiles.
\item The existence of the fundamental plane for giant ellipticals
and observed deviations of dwarf spheroidal galaxies from it
follow naturally from the dynamical equilibrium condition in the 
framework of the two-component model.
\item The major difference in the empirical luminosity--surface-brightness
relation for dwarf and giant families of galaxies could be explained
in the context of a smooth transition from dark matter dominated 
dwarfs to luminous matter dominated (in the centre) giants.
\item The comparison of the model predictions with empirical scaling
relations allows one to suggest that a common dissipative process operates
during the initial violent stage of development of stellar systems in
massive dark haloes of dwarf and giant galaxies. The outcome of the
initial burst of star formation is regulated by the depth of the 
potential well of the galaxy halo. 
\end{enumerate}
These results should be considered as preliminary and indicating the
direction for more detailed theoretical work. Also, new observations are
necessary to define more exactly the scaling relations for dwarf
galaxies. 

\section*{Acknowledgments}

This work was partly supported by the Russian
Foundation for Basic Research (project code 93-02-02957).
The author is grateful to Bruno Binggeli, Massimo Persic and Massimo
Stiavelli for useful comments on the manuscript, and to the staff of MPA for
their warm hospitality.

\bibliography{../Bibliography/cflow,../Bibliography/ellip}

\bsp
\label{lastpage}

\end{document}